# RING FOR TEST OF NONLINEAR INTEGRABLE OPTICS

A. Valishev, S. Nagaitsev, V. Kashikhin FNAL, Batavia, IL 60510, U.S.A.
V. Danilov SNS, Oak Ridge, TN 37830, U.S.A.

*Abstract*

Nonlinear optics is a promising idea potentially opening the path towards achieving super high beam intensities in circular accelerators. Creation of a tune spread reaching 50% of the betatron tune would provide strong Landau damping and make the beam immune to instabilities. Recent theoretical work [1] has identified a possible way to implement stable nonlinear optics by incorporating nonlinear focusing elements into a specially designed machine lattice. In this report we propose the design of a test accelerator for a proof-of-principle experiment. We discuss possible studies at the machine, requirements on the optics stability and sensitivity to imperfections.

## INTRODUCTION

The theory of nonlinear integrable beam optics was discussed in detail in [1,2]. This work demonstrated that using conventional and special nonlinear focusing magnets it is possible to construct an accelerator lattice, in which the betatron motion is completely nonlinear yet stable. The strong nonlinearity of betatron motion would result in a significant spread of betatron tunes of particles within the bunch (up to 50% of the nominal tune), thus providing strong Landau damping of coherent instabilities.

The lattice cell (element of periodicity) of such a machine consists of two main parts:
1. Linear focusing (and possibly bending) block equivalent to an axially symmetric lens, with the betatron phase advance in both planes equal to a multiple of $\pi$.
2. Special nonlinear magnet placed in a straight section with equal beta-functions.

The machine may be comprised of an arbitrary number of cells.

Because of the simple requirements, the focusing block can be built using conventional dipole and quadrupole magnets, and house the necessary accelerator components, such as beam instrumentation, injection and RF system, etc.

The superconducting RF test facility, currently under construction at the New Muon Lab (NML) at Fermilab [4], will provide electron beam with the energy up to 750 MeV for the needs of experiments on advanced accelerator R&D. The experimental hall at NML provides space sufficient for a test storage ring, which could be used for a proof-of-principle experiment on the nonlinear integrable beam optics.

In this report we discuss the choice of parameters and describe the design of the storage ring and the nonlinear magnet. We present the results of numerical simulations of the particle motion in the test accelerator and study the sensitivity of the design to imperfections.

## MACHINE LATTICE

In our design the ring is made of four cells. The cell consists of two mirror symmetric halves, each with 5 quadrupoles and one dipole bending by 45 degrees. With the betatron phase advance per cell of 0.8 (in units of $2\pi$), this makes the total betatron tune of 3.2. Hence, in the extreme case the maximum tune shift generated by the nonlinear magnets may reach 1.6, meaning some particles in the bunch will cross an integer resonance.

This design also provides 2 m insertions for the nonlinear magnets and four 1 m long straight sections for injection, RF and other systems. The minimal number of quadrupole magnets required to implement an axially symmetric lens in a straight section is five. The large number of quadrupoles we use permits a wide range of tuning for the betatron tune, which can be varied between 2.4 and 3.6, and dispersion and momentum compaction.

Table 1 lists the main parameters of the machine. Figures 2,3 show the machine layout in the NML experimental hall, and lattice functions of one cell, respectively.

Table 1: Machine parameters

| Parameter | Value |
| --- | --- |
| Beam energy | 150 MeV |
| Circumference | 32 m |
| Bending dipole field | 0.5 T |
| RF voltage | 50 kV |
| Maximum $\beta$-function | 7 m |
| Momentum compaction | 0.124 |
| Betatron tune | $Q_x, Q_y$=3.2 (2.4 to 3.6) |
| Equilibrium transverse emittance r.m.s. non-normalized | 0.06 $\mu$m |
| Synchrotron radiation damping time | 1 s ($\sim 10^7$ turns) |

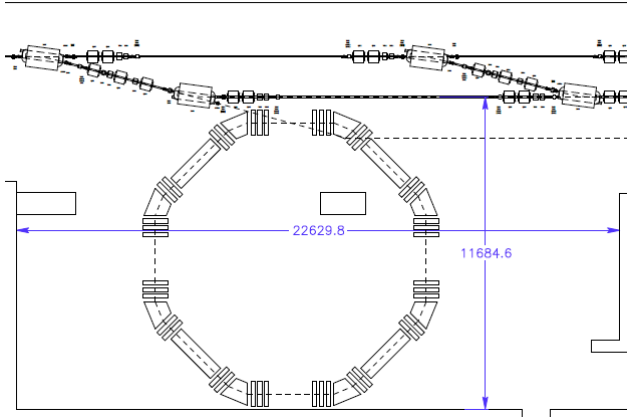

Figure 1: Machine layout.

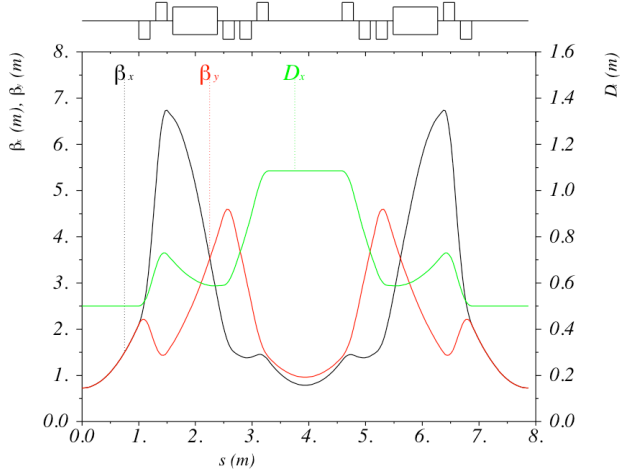

Figure 2: Lattice functions of a cell.

## NONLINEAR MAGNET

The condition of the Hamiltonian time-independence requires that the nonlinear potential must *continuously* change along the length of the nonlinear section (eq. 4 in [2]). The potential is defined by two parameters: the strength parameter $t$, and the geometric parameter $c$, which represents the distance between the singularities, or the element aperture (see eq. 10 in [2]). The geometric parameter $c$ scales as the square root of $\beta$-function in the nonlinear straight section, and the strength parameter $t$ scales as $1/\beta$. Since it's not practical to manufacture a magnet with complex varying aperture, we consider approximating the continuously varying potential with a number of thin magnets of constant aperture. Figure 3 shows the distribution of strength of the lowest (quadrupole) harmonic in 20 thin magnets.

The magnetic potential can be expanded into multipole series

$$U(x,y) = -\frac{t}{c^2}\text{Im}\left((x+iy)^2 + \frac{2}{3c^2}(x+iy)^4 + \frac{8}{15c^4}(x+iy)^6 + ...\right).$$

However, this expansion is only valid inside the $r = \sqrt{x^2+y^2} < c$ circle. Vertical oscillation amplitudes $y>c$ are essential for achievement of large tune spread. By proper shaping of the magnetic poles we were able to achieve good field quality in the region $x<c$, $y<2\times c$ (Fig. 4.). The magnet can be optimized further to include fringe-field effects and extend the good field region.

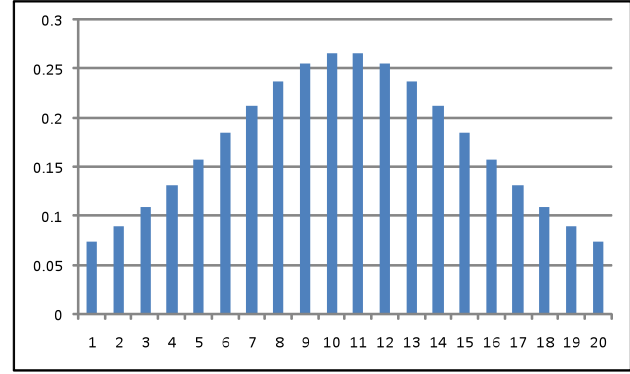

Figure 3: Distribution of quadrupole component (T/m) in the nonlinear magnet section.

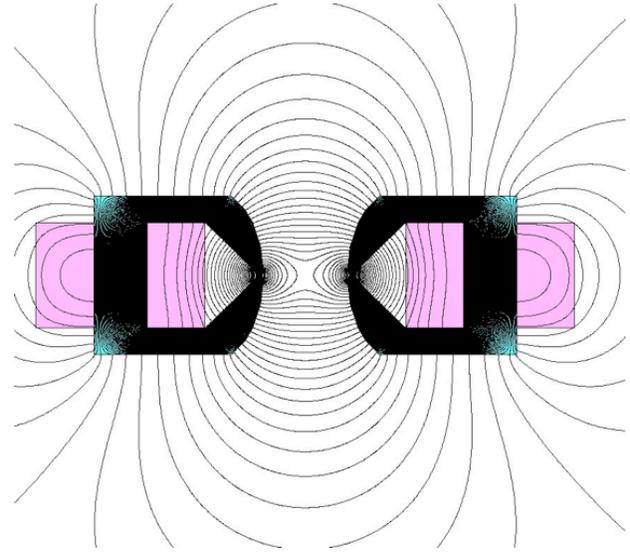

Figure 4: Cross-section of nonlinear magnet.

## NUMERICAL SIMULATIONS

We used numerical particle tracking to determine the effect of the nonlinear machine realization features on the preservation of particle motion stability, and to define the possible experiments. For this the nonlinear lens was implemented in MAD-X [4] tracking module as a thin element.

We evaluated the possibility of using a truncated multipole expansion of the potential instead of the true nonlinear potential. Since the expansion is valid inside the $r<c$ aperture, this limits the maximum particle amplitudes to approx. $r<0.5\times c$. Figure 5 demonstrates that the multipole expansion with 5 components (up to n=9 in field) works well at these amplitudes. In this case it is possible to achieve the vertical tune spread of 18% (0.15 per cell with the phase advance $Q_0$=0.8, or 0.45 for the entire machine). This approach offers the advantage of using conventional magnet technology and the convenience of round beam pipe aperture but does not allow extremely high tune spreads.

To achieve the tune spread exceeding 20% it is necessary to allow large vertical amplitudes. This defeats the validity of multipole expansion, and a true nonlinear element must be used. Figure 6 shows the tune spread per one cell for the case of the beam contained within an elliptical aperture with the ratio of horizontal/vertical dimensions equal to 7. The strength parameter $t<0.5$ keeps the motion in the vicinity of 0 stable.

When $t$ exceeds 0.5 (Fig. 7), it is still possible to contain the beam within the elliptical aperture. The vertical tune spread per cell in this case is 0.5. For a machine consisting of four cells, the total tune spread is 1, and some particles cross the $Qy=3$ integer resonance.

Simulations have shown that approximation of the nonlinear potential with 20 evenly spaced thin elements does not influence stability for the tested range of parameters. The tolerances on the optics of the focusing block are achievable with the conventional methods of linear optics tuning. Specifically, the deviation of the phase advance from 0.5 can reach 0.005, and the beta-functions in the nonlinear magnet section can be different by 10% without adverse effects. Synchrotron oscillations with the momentum deviation of 0.001 and uncorrected natural chromaticity of the lattice equal to -15 do not affect the dynamical aperture. Transverse misalignments of the individual nonlinear lenses with the r.m.s. of 0.5 mm are tolerable.

Synchrotron radiation damping determines the equilibrium beams size, which for the beam energy of 150 MeV is 0.2 mm at the middle of nonlinear section. Thus, a significant tune spread can be obtained for comparable nonlinear magnet dimensional parameter $c$ (aperture). Operating a machine with such beam pipe aperture is impractical. Instead, we propose to allow a more realistic $c\sim1$ cm, and then sample the aperture of the machine with the small emittance beam delivered by the linac. The synchrotron radiation damping time of $10^7$ turns allows ample time for the tune measurement. Alternatively, lower beam energy may provide a more practical ratio between the equilibrium beam size and the magnet aperture.

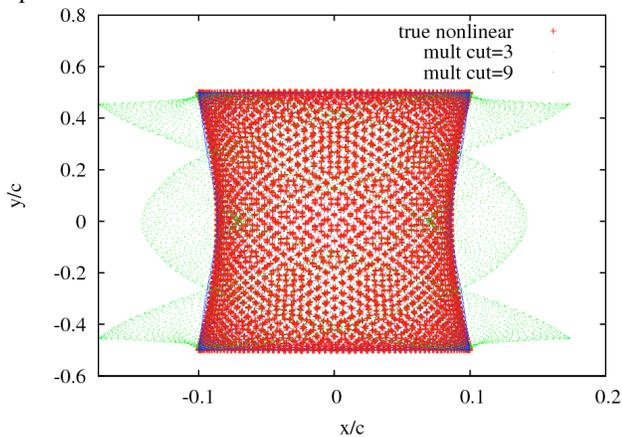

Figure 5: Comparison of tracking with true nonlinear potential (red dots) and multipole expansion truncated at octupole n=3 (green dots), n=9 (blue dots).

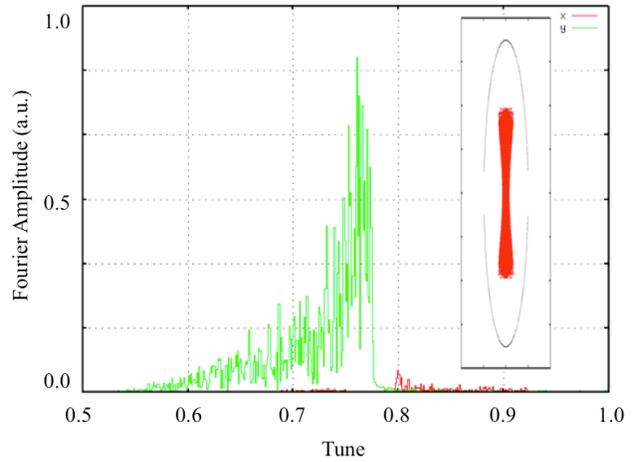

Figure 6: Spectrum of dipole moment for tracking in one cell with unperturbed tune $Q_0$=0.8, $t$=0.45. Inset: beam in the aperture (A$x$=10 mm, A$y$=70 mm)

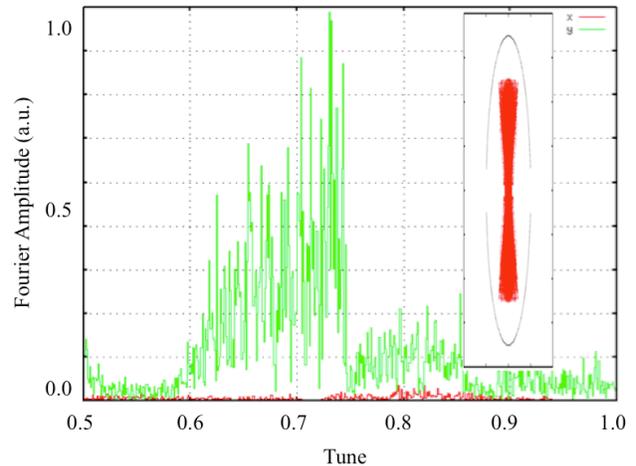

Figure 7: Spectrum of dipole moment for tracking in one cell with unperturbed tune $Q_0$=0.8, $t$=1.5. Inset: beam in the aperture (A$x$=10 mm, A$y$=70 mm)

## ACKNOWLEDGMENTS

We would like to thank F. Schmidt (CERN) and E. Forest (KEK) for their help in implementation of the nonlinear element in MAD-X and PTC codes.

## REFERENCES

[1] V. Danilov and S. Nagaitsev, Phys. Rev. ST Accel. Beams 13, 084002 (2010)
[2] S. Nagaitsev, V. Danilov, "A Search for Integrable Four-Dimensional Nonlinear Accelerator Lattices", in Proceedings of IPAC'10, Kyoto, Japan 2010. THPE094
[3] M. Church et al., "Plans for a 750 MeV Electron Beam Test Facility at Fermilab", in Proceedings of PAC07, Albuquerque, NM 2007. THPMN099
[4] F. Schmidt, "Update on MAD-X and Future Plans", in Proceedings of ICAP09, San Francisco, CA 2009. WE3IOPK04